\documentclass[aps,pra,twocolumn,showpacs,floatfix]{revtex4}
\usepackage{amsmath}
\usepackage{epsfig}
\usepackage{graphicx}
 
\newcommand{\ket}[1]{\left. \vrule\; #1 \right\rangle}
\newcommand{\braket}[3]{\left\langle #1\;\vrule\; #2 \;\vrule\; #3 \right\rangle}

\newcommand{\scalprod}[2]{\left\langle #1\;\vrule\; #2 \right\rangle}
\newcommand{\cre}[2]{\hat{#1}^\dag_{#2}}            
\newcommand{\ann}[2]{\hat{#1}_{#2}}                 

\newcommand{\dblint}{\int \!\!\! \int}         
\newcommand{\abs}[1]{\vrule\,#1\,\vrule\,}      
\newcommand{\p}{^\prime}          

\newcommand{\vect}[1]{\mathbf{#1}}       

\newcommand{\rr}{{\vect{r}}}

\begin{document}

\title{Universality of Many-Body States in Rotating Bose and Fermi Systems}

\author{M.~Borgh$^1$, M.~Koskinen$^2$, J.~Christensson$^1$, M.~Manninen$^2$,
  and S.~M.~Reimann$^1$}  
\affiliation{$\mbox{}^1$Mathematical Physics, LTH, Lund University,
22100 Lund, Sweden\\
$\mbox{}^2$NanoScience Center, 40014 University of Jyv\"{a}skyl\"{a}, Finland}
\date{\today}

\begin{abstract}
We propose a universal transformation from a many-boson state to a
corresponding many-fermion state in the lowest Landau level approximation of
rotating many-body systems, inspired by the Laughlin wave function and by the
Jain composite-fermion construction. We employ the exact-diagonalization
technique for finding the many-body states. The overlap between the
transformed  boson ground state and the true fermion ground state is
calculated in order to measure the quality of the transformation. For very
small and high angular momenta, the overlap is typically above 90\%. For
intermediate angular momenta, mixing between states complicates the picture
and leads to small ground-state overlaps at some angular momenta.
\end{abstract}

\pacs{60.10.-j, 70.10.Pm, 68.65.-k, 03.75.Kk, 32.80.Pj}

\maketitle

The properties of rotating many-body systems have been the topic of intense
study, theoretical as well as experimental~\cite{dalfovo1999,butts1999,
  mottelson1999,bertsch1999,matthews1999,kavoulakis2000,romanovsky2004,
  toreblad2004, jeon2004,
  chang2005a,chang2005b,
  cazalilla2005,jeon2005,barberan2006,reimann2006,
  regnault2006,yannouleas2006,jeon2007,shi2007,dai2007}.  
In particular, Bose-Einstein condensates have been of
great interest since the advent of the laser-cooling
technique~\cite{cohen-tannoudji1998}. It is 
by now a well-established fact that quantum many-body systems under rotation
form quantized vortices, a property that has been known in the contexts of
superconductivity and superfluidity~\cite{degennes,abrikosov1957}. In earlier
works, the formation of vortices in few-body systems was studied quite
extensively, including both boson and fermion
systems~\cite{toreblad2004,reimann2006,bargi2006}, as well as
their two-component
generalizations~\cite{bargi2006,bargi2007,koskinen2007}. In
Ref.~\cite{toreblad2004} we noted that some  
properties of vortex formation in few-body systems of repulsively interacting
particles in the lowest Landau level are universal, not only with respect to
the details of the 
interaction (the boson system was studied both with the long-range Coulomb
interaction and with a short-range, contact interaction), but also with
respect to the statistics of the constituent particles. We found remarkable
similarities between the yrast spectra of the boson and fermion systems, and
we also noted that vortices enter at very specific, and corresponding, angular
momenta in the boson and fermion systems, respectively. 

In the present
paper, we formalize this boson--fermion universality by means of a
transformation from a bosonic many-body wave function to a corresponding
fermionic wave function. The transformation is inspired by the Laughlin wave
function~\cite{laughlin1983}, and forms a direct parallel to the Chern--Simons
transformation in the theory of the fractional quantum Hall
effect~\cite{zhang1989,murthy2003} and Jain's composite-fermion 
picture~\cite{jain1989}. (The latter was recently successfully applied to the
problem of small quantum dots in magnetic fields, and other small quantum
systems at high angular
momentum~\cite{jeon2004,chang2005a,chang2005b,cazalilla2005,
  jeon2005,regnault2006, jeon2007,shi2007}.) 
This allows us to investigate quantitatively to what
extent this universality holds in a general comparison between few-body boson
and fermion systems in the lowest Landau level. The particles may be electrons
in a quantum dot (in the case of fermions) or optically or magnetically
trapped ions at ultra-low temperatures. In this work we concentrate on
harmonically confined, Coulomb-interacting particles in two dimensions.

This paper is organized as follows: The next section gives an introduction
to the mathematics of the proposed boson--fermion universality of many-body
systems in the lowest Landau level. This section also introduces the concepts
used throughout the paper, and defines a mathematical transformation of a
many-boson wave function into a corresponding many-fermion wave
function. Section~\ref{sec:calculations} gives a brief description of the
exact-diagonalization method in the lowest Landau level, and of the
implementation of the boson--fermion transformation. The reader who is
familiar with exact diagonalization may skip directly to
section~\ref{sec:results}, where our results are presented and discussed. Some
concluding remarks are given in section~\ref{sec:conclusions}. 

\section{Introduction to Boson--Fermion Universality}
\label{sec:intro}

In order to formulate a mathematical expression for the direct
comparison between boson and fermion many-body wave functions, we
use complex coordinates for the two-dimensional plane, $z=x+iy$.
The general wave function in the lowest Landau
level of one particle in a harmonic-oscillator potential is $\psi_\ell(z)
\propto z^\ell e^{-|z|^2}$, where $\ell$ is the angular momentum. (Atomic
units and one fixed frequency $\omega=1$ of the harmonic confinement are used
throughout the paper.)
In the boson system at zero angular momentum, all particles
reside in the $\ell=0$ state, and the many-body wave function is  
\begin{equation}
\label{eq:bosecond}
   \Psi^B_0 \propto e^{-\sum_k|z_k|^2}.
\end{equation}
As the system is set rotating, particles are lifted from the $\ell=0$
single-particle state in order to carry angular momentum. A single vortex is
formed at the center of mass when the total angular momentum of the system
equals the 
number of particles, in which case all particles are in the $\ell=1$
state. As angular momentum increases further, more vortices successively enter
the system. As detailed in Ref.~\cite{toreblad2004}, a vortex-generating state carrying $n$ vortices can be obtained
from Eq.~(\ref{eq:bosecond}) by successive multiplications with symmetric
polynomials: 
\begin{eqnarray}
\label{eq:nvortices}
\Psi_{n} &=&\prod_{j_1}^N (z_{j_1}-ae^{i\alpha_1})\times \cdots\times
\prod_{j_n}^N (z_{j_n}-ae^{i\alpha_n}) \Psi^{B}_0 \nonumber\\
&=& \prod_j^N (z_j^n-a^n)\Psi^{B}_0.
\end{eqnarray}
This wave function describes a state with $n$ vortices evenly spaced on a ring
with radius $a$ centered at the origin, and may be used to obtain a trial
many-body wave function~\cite{toreblad2004}.

The same line of reasoning can be repeated completely analogously for
fermions. However, in the lowest Landau level the angular
momentum of a many-fermion system cannot be zero. Instead the smallest
possible total angular momentum is achieved by putting one fermion in each of
the $N$ lowest single-particle states with single-particle angular momenta
ranging from $0$ to $N-1$. This state is called the {\em maximum density
  droplet\/} (MDD) and is the fermion equivalent of the zero-angular-momentum
states for bosons. The wave function of the MDD is given
by the Laughlin wave function with filling factor one~\cite{macdonald1993}:
\begin{equation}
\label{eq:MDD}
   \Psi^F_{\rm MDD} \propto \prod_{i<j}^N (z_i-z_j)e^{-\sum_k|z_k|^2}.
\end{equation}
In the construction of the vortex-generating state for the fermion system, the
boson condensate is replaced by the MDD.

Now we compare Eq.~(\ref{eq:bosecond}) with Eq.~(\ref{eq:MDD}) and note that
the latter can be obtained from the former simply by multiplication with the
polynomial 
\begin{equation}
\label{eq:fermpoly}
   D^F = \prod_{i<j}^N (z_i-z_j).
\end{equation}
Making the assumption that the transformation between the lowest angular
momentum states, $\Psi^F_{\rm MDD}=D^F\Psi^B_0$, holds whenever $\Psi^B_0$ is
replaced by $\Psi^F_{\rm MDD}$, we arrive
at a very general transformation from any bosonic many-body wave function to a
corresponding fermionic wave function, shifted in angular momentum by exactly
$L_{\rm MDD}$:
\begin{equation}
\label{eq:transformation}
  \Psi^F_{L_{\rm MDD}+L_B} = D^F\Psi^B_{L_B}.
\end{equation}

The degree of universality of the structure of the many-body wave function
between boson and fermion systems will be reflected in how well the
transformation~(\ref{eq:transformation}) reproduces the true fermion wave
function. The straightforward way of determining this is to calculate the
wave function of a given many-fermion system both directly, in order to obtain
the true wave function, and by transforming the corresponding boson state
according to Eq.~(\ref{eq:transformation}). The overlap between the true wave
function and the transformed boson wave function can the be calculated:
\begin{equation}
\label{eq:overlap}
  O = \;\abs{\!\!\scalprod{\Psi^F_{L_{\rm MDD}+L_B}}{D^F\Psi^B_{L_B}}}^2.
\end{equation}

The argument given above for the boson--fermion correspondence is entirely
heuristic: it is based on observing the common features of fermion and boson
wave functions in the lowest Landau level. The correspondence formalized in
Eq.~(\ref{eq:transformation}) is ultimately justified by the explicit
demonstration of its performance in section~\ref{sec:results}. However, it is
reasonable to anticipate the existence of a boson--fermion transformation of
this type from boson-Chern--Simons theory~\cite{zhang1989,murthy2003}
and composite-fermion theory~\cite{jain1989,murthy2003}. A
fermionization of repulsively interacting bosons in the lowest Landau level
has been described by several authors~\cite{romanovsky2004, jeon2005,
  cazalilla2005, chang2005b, regnault2006, yannouleas2006}. In particular,
Cazalilla {\em et al.}~\cite{cazalilla2005}, Chang {\em et
  al.}~\cite{chang2005b} as well as Regnault {\em et al.}~\cite{regnault2006}
explicitly map interacting bosons onto 
non-interacting, spinless fermions by means of a composite-fermion
construction. It is known that the mean-field approximation of non-interacting
composite fermions loses accuracy as angular momentum increases, and effects
of residual interactions become important~\cite{jeon2004, chang2005b,
  regnault2006, jeon2007}. In the 
case of a harmonic interaction, the transformation~(\ref{eq:transformation})
between interacting bosons and interacting fermions was rigorously derived
recently by Ruuska and Manninen~\cite{ruuska2005}.

In the Chern--Simons approach to the fractional quantum Hall effect, the
many-body wave function $\Psi_e$ of the electrons of the two-dimensional
electron 
liquid is transformed into a wave function $\Psi_{\rm CS}$ of particles moving
in an effective 
magnetic field. The transformation is defined as follows~\cite{murthy2003}:
\begin{equation}
\label{eq:CS}
  \Psi_e=\prod_{i<j}\left[\frac{z_i-z_j}{\abs{z_i-z_j}}\right]^p\Psi_{\rm CS}.
\end{equation}
The prefactor determines the number of flux quanta associated with each
particle, and it is symmetric or anti-symmetric with respect to particle
interchange depending on whether $p$ is chosen even or odd. In the simplest
possible model, we choose $p=1$. Since the original
electron wave function is fermionic, this means that the Chern--Simons wave
function describes a corresponding boson system. The transformation between
the boson system and the fermion system is the given by the factor
$\prod_{i<j}\left[(z_i-z_j)/\abs{z_i-z_j}\right]$.

Jain took the Chern--Simons approach further to a more sophisticated ansatz in
the lowest Landau level by introducing his composite-fermion
theory~\cite{jain1989,murthy2003}. In this approach, the power
$p$ of the Chern--Simons factor is explicitly taken to be an even integer,
thereby making the composite particles fermions. Also, the prefactor is simply
taken to be the {\em Jastrow factor\/} $\prod_{i<j}(z_i-z_j)^p$, which
attaches vortices rather than flux tubes to the
particles~\cite{murthy2003}. The wave function of the original fermions is
retained by projection to the lowest Landau level:
\begin{equation}
\label{eq:CF}
  \Psi_e={\cal P}\prod_{i<j}\left(z_i-z_j\right)^p\Psi_{\rm CF},
\end{equation}
where ${\cal P}$ is the projection operator into the lowest Landau level.

Jain's construction and the bosonic Chern--Simons wave function immediately
suggest a mapping between a boson system and an equivalent fermion system at
some other magnetic field (or, equivalently, angular momentum) of the form
described by Eq.~(\ref{eq:transformation}). For a harmonic particle--particle
interaction, this relation is exact, and can be derived analytically by
constructing a spectrum-generating algebra, and applying the corresponding
ladder operators to the ground-state wave functions $\Psi_B=1$ and
$\Psi_F=\prod_{i<j}(z_i-z_j)$ respectively~\cite{ruuska2005}.

The numerical study presented in this paper shows that the transformation is
very general and yields a good approximation for the fermion wave function in
the lowest Landau level, also for Coulomb-interacting particles.

\section{The Exact-Diagonalization Method and the Boson--Fermion
  Transformation}
\label{sec:calculations}

In order to calculate the overlap, Eq.~(\ref{eq:overlap}), the many-body wave
functions must be obtained with sufficient accuracy. In this study we take as
our model system harmonically confined, Coulomb-interacting
particles, and we solve both the fermion and the boson problems with the
exact-diagonalization method. The boson wave function is transformed into a 
corresponding fermionic many-body wave function by means of
Eq.~(\ref{eq:transformation}), which we have implemented numerically to
operate on a bosonic wave function in the Fock--Darwin basis. 

Since the model system used in this study is that of a harmonic confining
potential, a natural choice of single-particle basis is the set of
eigenfunctions of the two-dimensional harmonic oscillator in the lowest Landau
level:
\begin{equation}
\label{eq:hocomplex}
   \psi_\ell(z) = A_\ell z^\ell e^{-|z|^2}.
\end{equation}
The many-body wave function is then built in the Fock--Darwin basis formed
from these single-particle states, and the Hamiltonian
(up to an additive constant) is straightforwardly written as
\begin{equation}
\label{eq:H}
   H = \sum_{i} \ell_i\cre{a}{i}\ann{a}{i} 
     + \frac{1}{2}\sum_{i,j,k,l} 
        U_{i,j,k,l}\cre{a}{i}\cre{a}{j}\ann{a}{k}\ann{a}{l},
\end{equation}
where the interaction matrix element is
\begin{equation}
\label{eq:Umatrixelement}
\begin{split}
   U_{i,j,k,l} &= \braket{ij}{\hat{U}(\rr,\rr\p)}{kl}\\
   &= \dblint \psi_i^\ast(\rr)\psi_j^\ast(\rr\p) U(\rr,\rr\p)
             \psi_k(\rr)\psi_l(\rr\p)\,d\rr\,d\rr\p.
\end{split}
\end{equation}
Since we are working with many-body systems at given total angular momentum
$L$, the one-body term yields the same energy contribution for all allowed
Fock states. Therefore we may drop this term, and diagonalize only the
interaction part of the Hamiltonian. Working in the
lowest Landau level already guarantees that we have a finite number of
possible Fock
states for any given total angular momentum, and in this study we have used
all of these in the basis. Thus no further
truncation of the Hilbert space is done. The resulting matrix is diagonalized
numerically, yielding the interaction-energy spectrum and the many-body
eigenstates. The diagonalization is performed using
the Lanczos algorithm~\cite{lanczos}, except at the smallest angular momenta,
where a full diagonalization is needed due to the small matrix sizes.

The exact-diagonalization method outlined above is applicable to both fermions
and bosons; only the commutation relations of the creation and annihilation
operators differ, and the many-body Fock--Darwin basis states must be given
the appropriate symmetry properties (Slater determinants for fermions, product
states for bosons).

The numerical implementation of the  transformation,
Eq.~(\ref{eq:transformation}), requires some care. A general bosonic many-body
wave function consists of a linear combination of a (possibly very large)
number of Fock states. Upon multiplication by the determinant $D^F$
(Eq.~(\ref{eq:fermpoly})), each bosonic Fock state is transformed into a
fermionic many-body wave function, which is in general not a basis state in
the fermion Fock--Darwin basis, but instead must be represented as a linear
combination of fermionic basis states. Thus, each Fock state in the boson wave
function will yield several Fock states in the final fermionic wave function,
and correspondingly, the weight of each Fock state in the final fermionic wave
function may receive contributions from several Fock states in the original
bosonic wave function. 
\begin{figure}[t]
\includegraphics[angle=0]{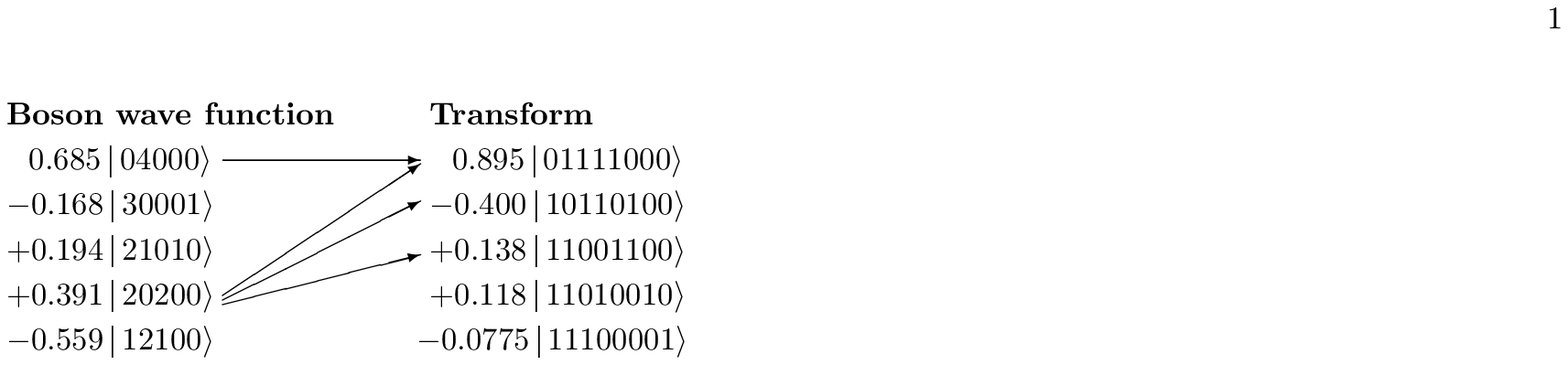}
\caption{As an example, we show the transformation of the $L_B=4$ boson wave
  function into a fermionic 
  wave function with $L_F=10$. One term in the boson wave
  function may contribute to the weight of several terms in the transformed
  wave function. Correspondingly, terms in the transformed
  wave function may receive contributions from more than one term in the
  original wave function.} 
\label{fig:transformation}
\end{figure}

For example, the four-particle bosonic state
$\ket{202}$ (with angular momentum $L_B=4$) transforms into the fermionic wave
function $0.79\ket{110011}-0.56\ket{101101}+0.25\ket{011110}$, a linear
combination of three states from the Fock basis (total angular momentum
$L_F=10$). The true wave function for the four-boson system with angular
momentum $L_B=4$ is a linear combination of five basis states, as depicted in
Fig.~\ref{fig:transformation}.
The state in our toy example appears in the fourth 
term. Under application of the transformation, Eq.~(\ref{eq:transformation}),
this wave function transforms into the fermionic wave function shown in the
right column of Fig.~\ref{fig:transformation}. As indicated by arrows, the
bosonic 
state $\ket{202}$ contributes to the first three terms of the transformed wave
function. Contributions come also from other terms in the original wave
function. For example, contributions to the first term come 
from all terms of the bosonic wave function. The contribution from the boson
basis state $\ket{04000}$ is also indicated in the figure (in fact, this
particular term in the boson wave function contributes only to the first term
of the transformed wave function). This example
shows that it is necessary to carefully collect the weights of the basis
states of the fermionic wave function resulting from the transformation.

\section{Results}
\label{sec:results}

We perform 
calculations of energies, wave functions and ground-state overlaps
systematically over a wide range of angular momenta for systems containing
between four and eight Coulomb-interacting particles. 

\begin{figure*}[tb]
   \includegraphics[angle=0,width=17cm]{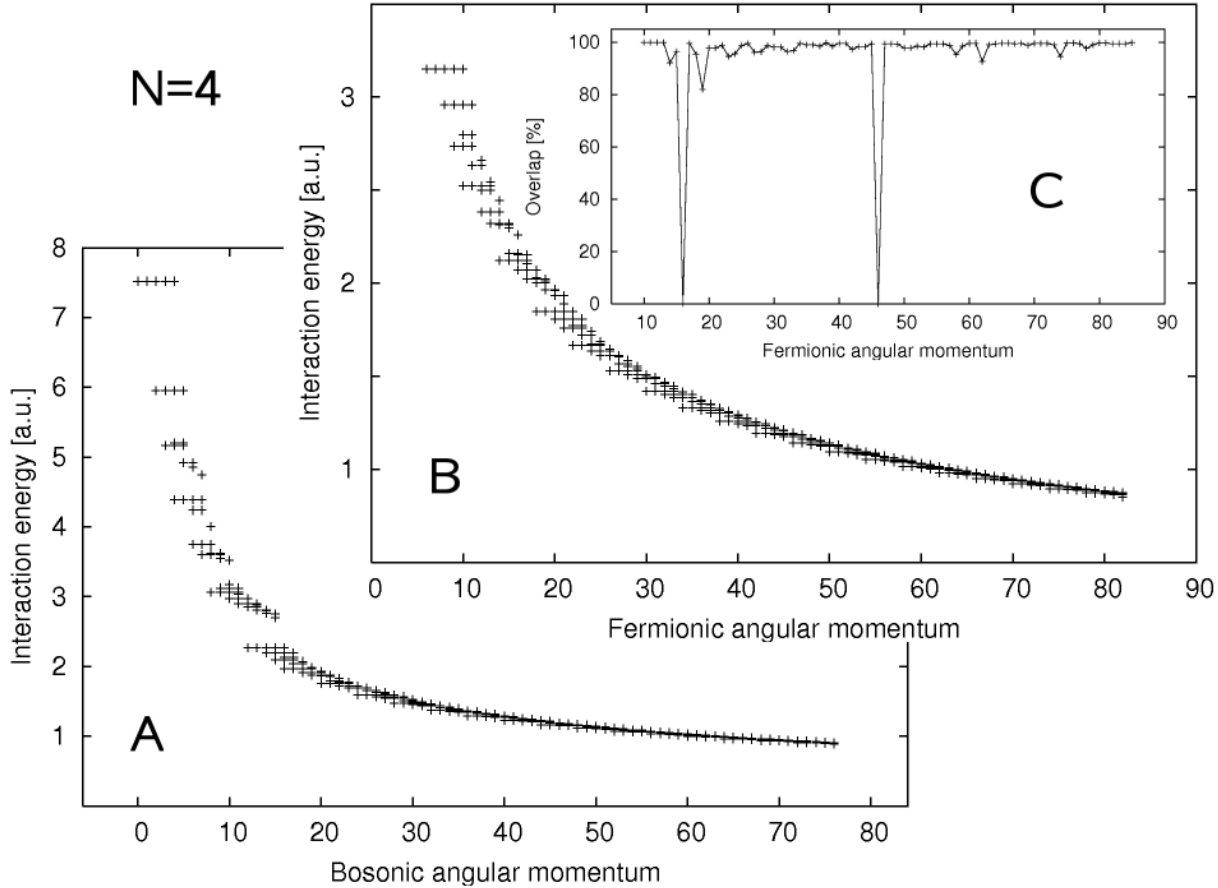}
\caption{Yrast spectra and overlaps for a four-particle system. {\bf (A)}
   Yrast spectrum of four harmonically confined bosons. The five
   lowest states for each angular momentum are included. {\bf (B)}
   The corresponding yrast spectrum for four fermions. {\bf (C)} Overlap of
   the transformed boson ground state with the fermion ground state,
   calculated from Eq.~(\ref{eq:overlap}).} 
\label{fig:n4ol}
\end{figure*}

Fig.~\ref{fig:n4ol} shows the overlap of the ground state of a
four-fermion system with the transformed ground state of a corresponding
four-boson system, for total fermion angular momenta in the range
$L_F=18$--$62$. In the small four-particle system, the overlap is very good,
close to 100\% for all angular momenta. The top and middle panels of the
figure show the yrast spectra of the five lowest states for the boson and
fermion systems respectively. The similarity of the yrast lines is readily
apparent from this figure. Both systems show a four-fold periodicity, and the
kinks in the yrast line---where there is also a relatively large gap between
the ground state and the first excited state---appear at corresponding
angular momenta, $L_F=L_B+L_{\rm MDD}$. The overlap oscillates with the same
four-fold period for high angular 
momenta, with the overlap having a local maximum whenever there is a kink in
the yrast line.

\begin{figure}[tb]
  \includegraphics[angle=0,width=8.8cm]{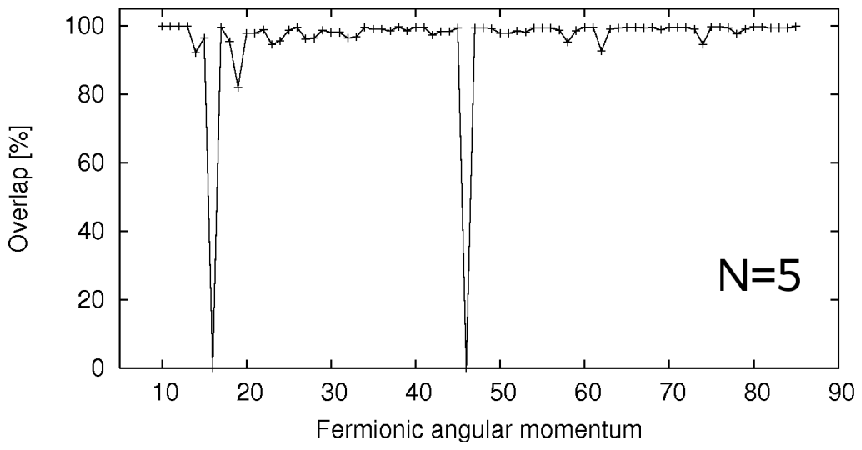}
  \includegraphics[angle=0,width=8.8cm]{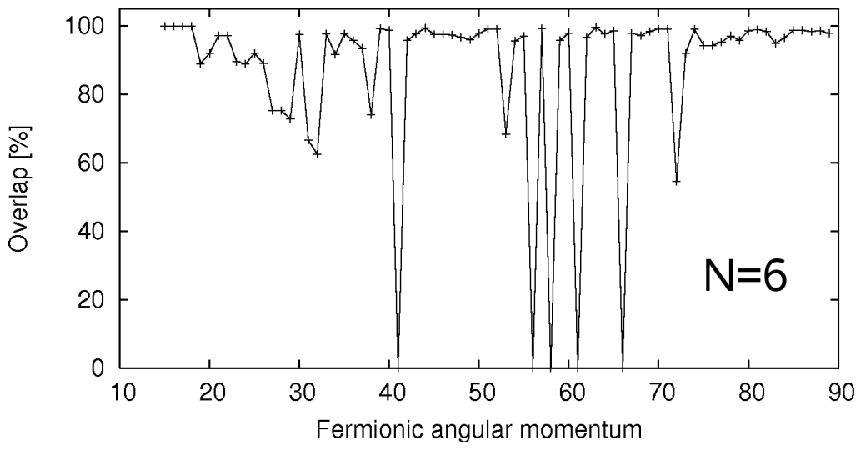}
\caption{Overlap between the true fermion ground state and the transformed
   boson ground state as a function of total angular momentum for five (top
   panel) and six (bottom panel) particles.} 
\label{fig:n5n6ol}
\end{figure}

Looking at a system with more particles, the picture becomes more
complicated. Fig.~\ref{fig:n5n6ol} shows the ground-state overlap in the five-
and six-particle systems over the entire range of total angular momenta
studied, $L_F=L_{\rm MDD}=10$ to $L_F=85$ for five particles and $L_F=L_{\rm
  MDD}=15$ to $L_F=89$ for six. At the smallest angular momenta, the overlap
is always 100\% due to the fact that the Hilbert space is so small that only
one or a few wave functions are possible. It is interesting to note that also
in the limit of high angular momentum, the overlap between the true fermion
ground state and the transformed boson ground state tends towards 100\%. This
shows that the performance of Eq.~(\ref{eq:transformation}) is not a
consequence of the Laughlin wave function being a good approximation. 
In fact, the overlap between the Laughlin wave function and the exact wave
function decreases with decreasing filling factor~\cite{yannouleas2003}.

Fig.~\ref{fig:n5n6ol} also shows some cases where the overlap drops to
zero. These cases will be discussed in some detail later in this section

Away from the extremes of very small and very large angular momenta, the
overlap drops and begins to fluctuate. In the case of five particles (top
panel of Fig.~\ref{fig:n5n6ol}) this is seen as a general decrease of the
overlap for angular momenta in the range $L_F=14$--$35$ together with two drops
to zero at $L_F=16$ and $L_F=46$. This pattern is repeated in a much more
complicated picture in the six-particle case, shown in Fig.~\ref{fig:n5n6ol},
bottom panel. Again, there is a general decrease in the overlap as the total
angular momentum increases beyond the small angular momenta where the overlap
is large due to the small Hilbert space. As the angular momentum increases
beyond $L_F\approx 40$ there is a region where the overlap tends to be either
large or vanishing. At the very largest angular momenta studied, the overlap
settles at over 90\%.

\begin{figure}[tb]
  \includegraphics[angle=0,width=8.8cm]{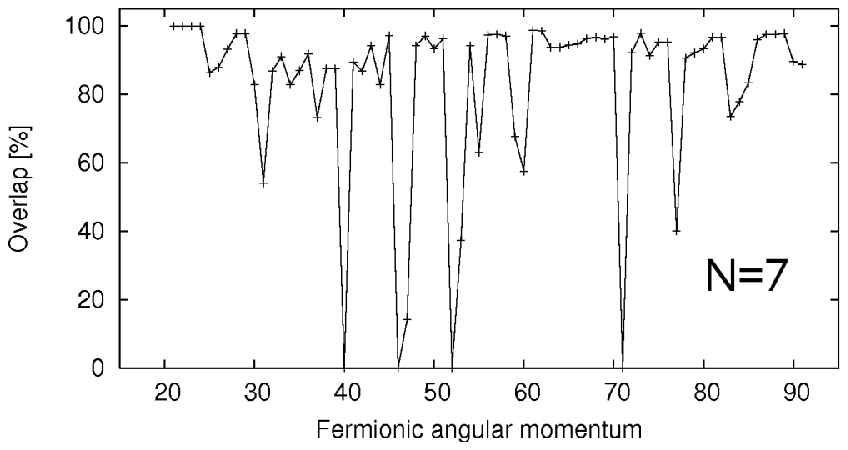}
  \includegraphics[angle=0,width=8.8cm]{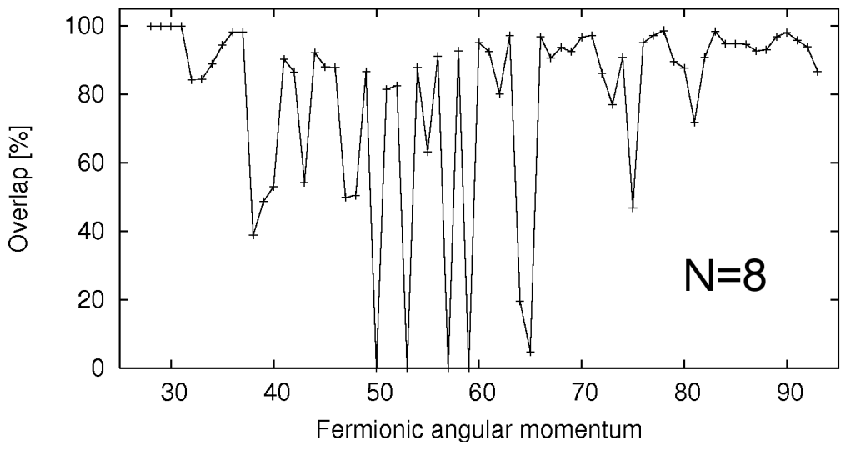}
\caption{Overlap between the true fermion ground state and the transformed
   boson ground state as a function of total angular momentum for seven (top
   panel) and eight (bottom panel) particles.}
\label{fig:n7n8ol}
\end{figure}

The behavior described for the five- and six-particle systems is echoed in
smaller and larger systems as well. In the four-particle system
(Fig.~\ref{fig:n4ol}) the overlap is
well above 90\% for all angular momenta studied, $L_F=6$--$82$. However, for
angular momenta in the range $L_F=10$--$40$ the overlap oscillates between
96\% and 99\%, and for $L_F=24$ dropping to 93\%, before returning to values
very close to 100\% for larger angular momenta. In systems with seven and
eight particles (Fig.~\ref{fig:n7n8ol}) the overlap oscillates wildly over
most of the studied region of angular momentum, where overlaps at or above
80\% are interspersed with drops to about 50\% or below. There is a tendency
towards better overlaps in the high angular momentum limit of the data.

We may conclude from these observations that the transformation defined in
Eq.~(\ref{eq:transformation}) from a bosonic many-body wave function to a
fermionic wave function for the same number of particles and angular momentum
$L_F=L_B+L_{\rm MDD}$ performs well in the low and high angular momentum
limits, where good performance may be expected due to the restricted Hilbert
space and particle localization respectively. For intermediate values of the
angular momentum, the performance of the transformation needs a deeper study.

The poor performance of Eq.~(\ref{eq:transformation}) for certain combinations
of particle number and angular momentum is a result of a
restructuring of levels in the yrast spectrum between the boson and fermion
systems. The result is that the boson ground state will have a different
structure than the fermion state, and therefore so will the transformed boson
wave function. In order to study the particle configurations corresponding to
the different many-body wave functions, we look at the {\em pair-correlation
  function\/} 
\begin{equation}
\label{eq.pc}
   g(\rr\p,\rr)
   = \braket{\Psi}{\cre{\psi}{}(\rr\p)\cre{\psi}{}(\rr)
   \ann{\psi}{}(\rr\p)\ann{\psi}{}(\rr)}{\Psi}, 
\end{equation}
which gives the probability of finding a particle at $\rr$, given that there
is a particle at $\rr\p$. 

\begin{figure}[tb]
   \includegraphics[angle=0,width=8cm]{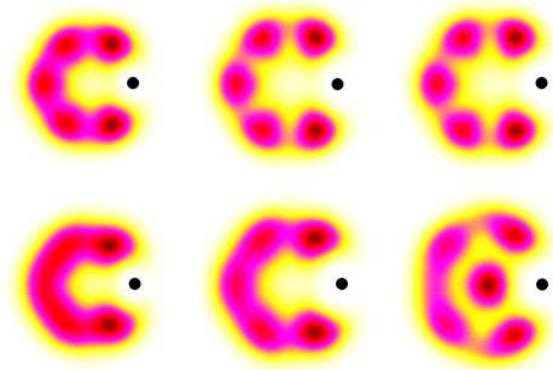}
\caption{Pair-correlation functions for the six-particle system at total
   angular momentum $L_F=57$, corresponding to $L_B=42$ in the boson system
   (top row), and at $L_F=58$ (bottom row). The panels in each row
   show, 
   from left to right, the pair-correlation function of the boson ground
   state, the transformed boson ground state, and the fermion ground state,
   respectively. The black dot indicates the position of the fixed reference
   particle}
\label{fig:n6lf57lf58pc}
\end{figure}

We may compare the pair-correlation functions
obtained from the true fermion wave function and from the transformed boson
wave function. Six particles at total fermionic angular momentum $L_F=57$ and
$L_F=58 $ are two examples from the intermediate angular momentum range where
the transformed 
boson ground state sometimes reproduces the fermion ground state very well,
but sometimes fails spectacularly. The former is the case at $L_F=57$, while
the latter is true for $L_F=58$. 
Fig.~\ref{fig:n6lf57lf58pc} shows the pair-correlation functions of
the boson wave function (top panel), the transformed boson wave function 
(middle panel), and the fermion wave function (bottom panel) for $L_F=57$ (left
column) and $L_F=58$ (right column). The 
reference particle in each plot is indicated by a black dot in the
figure. Its position is chosen such that it sits at the distance $r_{\rm max}$
from the origin where the particle density has its maximum.

At $L_F=57$ the particle structure is extremely 
well reproduced by the transformation from the bosonic state with
$L_B=L_F-L_{\rm MDD}=42$, and the two pair-correlation functions from the true
fermion wave function and the transformed boson wave function are almost
indistinguishable. The true fermionic many-body wave function displays an
internal structure of localized particles, where the six particles of the
system are equidistantly situated on a circle. This is one of two classically
\mbox{(meta-)stable} configurations for six coulomb-interacting particles in a
harmonic confining potential~\cite{bolton1993}. This structure is
also seen in the the bosonic wave function, and the effect of multiplication
with $D^F,$ apart from changing the particle-exchange symmetry, is to push the
particles outward.

The other classically stable configuration of six equal electrical
charges is 
a ring of five particles with the sixth particle sitting precisely at the
center of the ring. In the classical system, this configuration actually has a
slightly lower energy than the ring of six particles~\cite{bolton1993}. These
two configurations, $(6,0)$ and $(5,1)$, are known to 
compete in quantum-mechanical six-body systems with repulsive coulomb
interaction~\cite{reimann2000,reusch2001}. 
In an electron system where the
spin degree of freedom is not 
frozen out, the $(6,0)$ configuration tends to be favored (unless the system
is very dilute), because of frustration in the $(5,1)$
configuration~\cite{reimann2000,reusch2001,ghosal2007}. In the 
present study, we deal with systems of spinless bosons and of spin-polarized
fermions. Therefore, spin frustration does not come into play, and the two
classical configurations compete. 
When the angular momentum is increased just
by one unit to $L_F=58$, the fermion ground state is the $(5,1)$
configuration, as shown in the bottom right panel of
Fig.~\ref{fig:n6lf57lf58pc}. In the boson system, however, another change
takes place with the corresponding increase of one unit of angular momentum:
the particles remain seated on one ring of six particles, but the degree of
localization decreases. Correspondingly, the multiplication
with $D^F$ expands 
the state to a ring of six slightly smeared out maxima, a state that is
orthogonal to the true ground state.

\begin{figure}[t]
\includegraphics[angle=0,width=8cm]{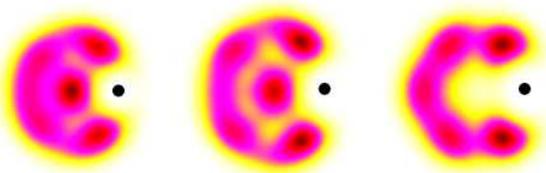}
\caption{Pair-correlation functions for the first excited state in the
   six-particle system at total 
   angular momentum $L_F=58$, corresponding to $L_B=43$ in the boson
   system. The panels show 
   the pair-correlation function of the boson wave
   function (left), the transformed boson wave function (center), and the
   fermion wave function (right), respectively. The black dot indicates the
   position of the fixed reference particle.}
\label{fig:n6lf58exc}
\end{figure}

Indeed the orthogonality of the transformed state to the true fermion ground
state is no coincidence, as in fact the transformed bosonic ground state
corresponds to the first excited state of the fermion system, and
correspondingly the fermion ground state is well reproduced by applying the
transformation to the first excited state of the boson
system. Table~\ref{tab:n6lf58ol} shows the overlap of the fermion ground-state
wave function with each of the transformed wave functions of the five lowest
boson states. The same table also shows the overlap of the transformed boson
ground-state wave function with each of the five lowest states of the fermion
system. As can be seen from the table, the transformed bosonic ground state
almost perfectly reproduces the first excited state of the fermion
system. Plotting the pair-correlation functions, as done in
Fig.~\ref{fig:n6lf58exc}, confirms this picture. This figure shows the
pair-correlation functions of the first excited state in the boson system and
the transformed version of this state, and the first excited fermionic
state. These plots may be compared with the corresponding ground-state pair
correlations in Fig.~\ref{fig:n6lf57lf58pc}.
\begin{table}[t]
\caption{\label{tab:n6lf58ol} Overlap of the fermion ground state with the
  transformed wave function of each of the five lowest states of the boson
  system (middle column), in the system with $N=6$ particles and fermionic
  angular momentum $L_F=58$. The table also shows the overlap between the
  transformed boson ground state with each of the five lowest fermion states
  in the same system (right column)}
\begin{center} 
\begin{tabular}{c|c|c}
$s$ & $\;\abs{\!\scalprod{\Psi^F_{\rm g.s.}}{\Psi^B_s}\!}^2$ [\%] &
$\;\abs{\!\scalprod{\Psi^F_s}{\Psi^B_{\rm g.s.}}\!}^2$ [\%]\\ 
\hline\hline
g.s. & 0.00 & 0.00 \\
1st & 96.94 & 99.31 \\
2nd & 0.46 & 0.00 \\
3rd & 0.00 & 0.00 \\ 
4th & 0.00 & 0.00 \\
\end{tabular}
\end{center}
\end{table}

\begin{figure}[t]
   \includegraphics[angle=0,width=8cm]{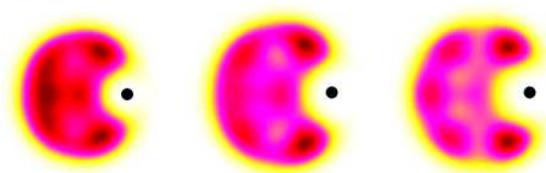}
\caption{Pair-correlation functions for the six-particle system at total
   angular momentum $L_F=53$, corresponding to $L_B=38$ in the boson
   system. The panels show the pair-correlation function of the boson ground
   state (left), the transformed boson ground state (center), and the fermion
   ground state (right), respectively. The black dot indicates the position
   of the fixed reference particle.}
\label{fig:n6lf53pc}
\end{figure}

$N=6$, $L_F=58$ discussed above is a particularly clean example where the
many-body configurations exchange places between the fermion and the boson
spectra. More generally, states may mix, such that one fermionic state is
reproduced by some linear combination of transformed boson states. Such mixing
of states between the bosonic and fermionic systems accounts for the cases
where the 
overlap between the ground states is neither close to 100\% nor vanishing. One
example is $N=6$ particles at angular momentum $L_F=53$, where the overlap
between the ground states is 68.32\%. In this case the fermionic ground state
has a considerable overlap with all of the three lowest bosonic states. This
is shown in Table~\ref{tab:n6lf53ol}. The pair-correlation functions are shown
in Fig.~\ref{fig:n6lf53pc}. The transformed boson ground state displays a
smeared-out 
$(5,1)$ configuration, while the fermion ground state is a $(6,0)$
configuration distorted such that there is also an increased particle density
at the origin.
\begin{table}[t]
\caption{\label{tab:n6lf53ol} Overlap of the fermion ground state with the
  transformed wave function of each of the five lowest states of the boson
  system (middle column), in the system with $N=6$ particles and fermionic
  angular momentum $L_F=53$. The table also shows the overlap between the
  transformed boson ground state with each of the five lowest fermion states
  in the same system (right column)}
\begin{center} 
\begin{tabular}{c|c|c}
$s$ & $\;\abs{\!\scalprod{\Psi^F_{\rm g.s.}}{\Psi^B_s}\!}^2$ [\%] &
$\;\abs{\!\scalprod{\Psi^F_s}{\Psi^B_{\rm g.s.}}\!}^2$ [\%]\\ 
\hline\hline
g.s. & 68.32 & 68.32 \\
1st & 25.74 & 20.71 \\
2nd & 4.67 & 9.29 \\
3rd & 0.00 & 0.00 \\ 
4th & 0.00 & 0.00 \\
\end{tabular}
\end{center}
\end{table}

The described mixing of the order of the states between bosonic and fermionic
system is seen also in the other cases were the 
overlap between the ground states is bad, also for other particle numbers than
six. There are 
many examples at all particle numbers larger than four. One particularly
drastic example for eight particles is $L_F=57$, where the fermion ground
state has 91.20\% overlap with the tranformed wave function of the first
excited boson state. However, the transformed bosonic ground state does not
have any appreciable overlap (1.94\%) with the first excited fermion
state. Instead it corresponds more closely to the second excited state (with
an overlap of 89.91\%).

\begin{figure}[htb]
\includegraphics[angle=-90,width=8.5cm]{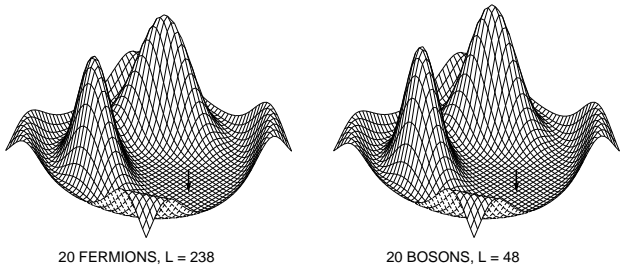}  
\caption{Hole--hole pair-correlation function describing the similar
vortex localization in boson and fermion systems. The boson wave function
was first converted to a fermion wave function as described in the text.}
\label{fig:holecorr}
\end{figure}

We have seen that when the particle number increases the overlap
between the fermion and boson states is still good, but in many
cases the lowest energy state of the fermion system corresponds to
an excited state of the boson system and vice versa. 
With increasing particle number the exchange of the order of levels 
becomes more important. This is already known in connection with the 
study of vortices in boson and fermion systems~\cite{reimann2006}.
For example, in the boson case the single vortex reaches the origin,
while in the fermion case the state with one vortex at the origin becomes an
excited state when the particle number is about 10 or greater. This is
also true for states with multiple vortices: the same vortex states are there
in both systems, but they appear as the lowest state at different (relative)
angular momenta. 
As an example, we may study  
three vortices in a rotating system of 20 particles.
For bosons the angular momentum is 48 and the state is the lowest-energy
state, while for fermions it is the fourth state 
at angular momentum 238. Figure~\ref{fig:holecorr} shows the 
{\em hole-hole correlation}~\cite{manninen2005} (i.e.\ pair correlation of
vortices) which can be calculated for
the fermionic wave functions (for the boson system, the transformation,
Eq.~\ref{eq:transformation}, is first applied to the wave function).
With this large number of particles we were not able to perform 
the computation with the full Fock-Darwin basis. The resulting overlap between
the converted boson and the fermion wave functions is only 88\%, but 
yet the resulting pair-correlation functions in Fig.~\ref{fig:holecorr} are
nearly indistinguishable. 

One may speculate that the large overlaps between fermionic wave function and
transformed bosonic wave function occur when the wave function is particularly
simple. One possible measure of the complexity of the fermion wave function
within the given single-particle basis (here the harmonic-oscillator
eigenstates) is the 
number of Fock states needed to make up 50\% of its norm. For six particles,
this number is five or below for angular momenta up to $L_F=30$, and the
overlap is 70\% and above, but we see very large overlaps also when the number
of Fock states needed for 50\% of the norm of the fermion wave function is
larger than 40. This shows that the good overlaps between fermion state and
transformed boson state are not merely the result of simple wave functions,
but that 
the transformation defined by Eq.~(\ref{eq:transformation}) can handle also
complex many body states.

\begin{figure}[htb]
  \epsfig{file=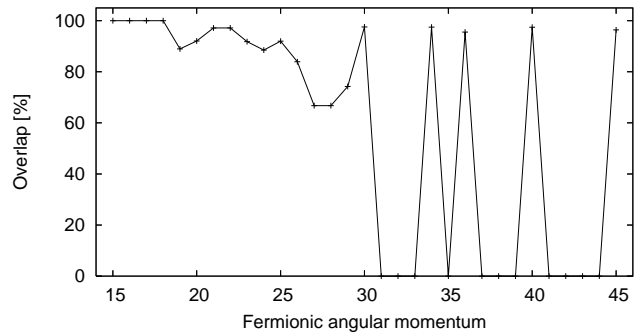,width=8.8cm,angle=0}
\caption{Overlap between the ground state of six Coulomb-interacting fermions
   and 
   the transformed ground state of six bosons with $\delta$-interaction,
   plotted as a function of total angular momentum in the fermion system.}
\label{fig:oldelta}
\end{figure}

The results detailed this far may to some extent be robust against changes
in the details of the repulsive interaction between the constituent particles
of the system. This is suggested by results where we calculate the overlap
between the ground-state wave function for Coulomb-interacting fermions and
the transformed ground-state wave function of bosons with a repulsive
$\delta$-type interaction. This is done for the case of six particles at
angular momenta between $L_F=15$ and $L_F=45$. The overlap is plotted as a
function of $L_F$ in Fig.~\ref{fig:oldelta}.
This figure should be compared with the same range of angular momentum in the
bottom panel of Fig.~\ref{fig:n5n6ol}. For the lower half of the range of
angular momenta, the overlap where $\delta$-interaction is used for the bosons
does not deviate substantially from the overlap where both systems have
Coulomb interaction. For these angular momenta, the greater part of the
many-body wave function comes from only a few Fock states and therefore we may
expect large overlaps regardless of the type of repulsive interaction, merely
due to the small Hilbert space. It is interesting to note, that not only are
the overlaps large (over 75\%), but the overlaps obtained from
$\delta$-interacting bosons follow those obtained from Coulomb-interacting
bosons closely.

These results should be expected from what we know about the Laughlin wave
function. The Laughlin wave function is a good approximation to the exact
many-body state as long as the angular momentum is not too large. Laughlin's
construction makes no explicit reference to the details of the repulsive
interaction between the constituent particles~\cite{laughlin1983}.

For angular momenta $L_F\gtrsim30$ the overlap drops to
zero except for certain angular momenta, where the overlap is very good (95\%
and above), and very close to the overlap obtained when Coulomb interaction is
used. In many cases the zero overlap is again simply the cause of different
orderings of the states in the the two systems, but in some cases also a
substantial mixing of states was found.

\section{Conclusion}
\label{sec:conclusions}

We conclude that for particles in the lowest Landau level there is a
far-reaching universality between bosons and fermions in the properties of the
rotating systems. This universality may be formulated mathematically as a
transformation (Eq.~(\ref{eq:transformation})) from a bosonic many-body wave
function to a fermionic 
one. These two wave functions will differ in total angular momentum exactly by
$L_{\rm MDD}$, the smallest possible angular momentum in the fermion
system. The transformation produces a very good correspondence (as measured by
calculating the overlap integral) between the bosonic and fermionic states
when the number of particles is small, when the angular momentum is very small
(due to the restricted Hilbert space), and when the angular momentum is large
(due to localization into states well described by Laughlin wave
functions). Away from these extremes, the correspondence between boson and
fermion states is more complicated. Apart from a general decrease of the the
overlaps due to the difficulty of reproducing the details of the more
complicated wave functions as the particle number increases, there is also
fluctuations between large and very small overlaps between the ground
states. This is due to a reshuffling of the levels in the spectrum between the
boson and fermion systems. In these cases it is not the ground state but a
low-lying excited state that transforms to the fermion ground states.

\begin{acknowledgments}
This work was financially supported by the Swedish Research Council, 
the Swedish Foundation for Strategic Research, the Academy of Finland, and
NordForsk.  
\end{acknowledgments}

\bibliography{borghref}  

\end{document}